\documentclass[structabstract]{aa}
\usepackage{graphicx}
\usepackage{txfonts}
\usepackage{epsfig}
\usepackage{natbib}
\usepackage[latin1]{inputenc}
\usepackage[dvips]{color}

\newcommand{\helixp}{\textsc{HeLIx$^\textbf{+}$}}
\newcommand{\hei}{He~\textsc{I}}

\begin{document}

\title{The Height of Chromospheric Loops in an Emerging Flux Region}
\subtitle{}

\author{L. Merenda\inst{1}, A. Lagg\inst{1}, S. K. Solanki\inst{1}\inst{2}
}

\offprints{L. Merenda}

\institute{Max-Planck-Institut f\"ur Sonnensystemforschung,
  Max-Planck-Stra\ss{}e 2, Katlenburg-Lindau, Germany\\
  \email{merenda@mps.mpg.de} \and School of Space Research, Kyung Hee University, Yongin, Gyeonggi, 446-701, Korea}

\date{Accepted for publication in Astronomy \& Astrophysics on June 1$^{st}$
  2011 }

\authorrunning{Merenda et al.}
\titlerunning{The Height of Chromospheric Loops}

\abstract
{The chromospheric layer observable with the \hei{} 10830 \AA\ triplet is
  strongly warped. The analysis of the magnetic morphology of this layer
  therefore requires a reliable technique to determine the height at which the
  \hei{} absorption takes place.}
{The \hei{} absorption signature connecting two pores of opposite polarity in
  an emerging flux region is investigated. This signature is suggestive of a
  loop system connecting the two pores. We aim to show that limits can be set
  on the height of this chromospheric loop system.}
{The increasing anisotropy in the illumination of a thin, magnetic structure
   intensifies the linear polarization signal observed in the
  \hei{} triplet with height. This signal is altered by the Hanle effect. We
  apply an inversion technique incorporating the joint action of the Hanle and
  Zeeman effects, with the absorption layer height being one of the free
  parameters.}
{The observed linear polarization signal can be explained only if the loop
  apex is higher than $\approx$5\,Mm. Best agreement with the observations is
  achieved for a height of 6.3\,Mm. }
{The strength of the linear polarization signal in the loop apex is
  inconsistent with the assumption of a \hei{} absorption layer at a constant
  height level. The determined height supports the earlier conclusion that dark He 10830 \AA\
  filaments in emerging flux regions trace emerging loops. }


\maketitle

\section{Introduction}

The magnetic field is commonly measured in the solar photosphere, while
measurements of the chromospheric and coronal field are rarer \citep[see,
e.g.,][for an overview]{solanki2006a}. Whereas photospheric magnetic field
measurements dominantly employ the Zeeman effect, in higher layers of the
solar atmosphere other methods come increasingly into play \citep[see][for a
review]{lagg2005}, in particular also the Hanle effect. One of the most used
diagnostics of the magnetic field in the upper chromosphere is the 10830 \AA\
\hei{} multiplet. For this multiplet the Zeeman effect is sensitive to
magnetic fields above $\approx 50$ Gauss, well suited to study magnetic fields
of active regions \citep[e.g.][]{harvey1971,rueedi1995,penn1995}. The Hanle
effect is sensitive to weaker fields (1-100 Gauss) and has been used to
measure the magnetic fields of prominences and spicules observed off-limb
\citep[e.g.][]{trujillobueno2005,merenda2006} and of filaments and emerging
flux regions observed on-disk \citep[e.g.][]{trujillobueno2005,lagg2004}.
Unlike the level of polarization produced by the Zeeman effect, the
polarization produced by scattering and then modified by the Hanle effect (in
both $90^{\circ}$ and forward scattering) depends on the anisotropy of the
radiation field exciting the scattering atoms.

This anisotropy increases with the height of the observed structure above the
solar surface \citep{Sahal-Brechot1977}. Therefore, this effect can in
principle be used to determine at what height in the chromosphere or corona
the observed magnetic structures are located. Since the \hei{} lines are
optically thin in some locations, notably in the quite Sun
\citep[e.g.][]{giovanelli1977}, but optically thick at others, in particular
in young active regions \citep{xu2009}, its formation could vary considerably,
making such an independent determination of the height an interesting addition
to the diagnostic capabilities of the He multiplet.

\cite{solanki2003} proposed an interpretation of spectropolarimetric
observations of an emerging flux region based on the assumption that the He
10830 \AA\ line is formed along the freshly emerged magnetic field lines,
which in this case reached $\approx 10$\,Mm. Recently, \citet{judge2009} has
proposed a different interpretation of these observations according to which
the magnetic field sampled by the observations is located in a thin layer at a
constant height ($\approx 2.4$\,Mm) near the top of the chromosphere. This
controversy lends an even stronger need for an independent diagnostic of the
height of the magnetic features sampled by the \hei{} 10830 triplet. In this
paper we develop the height diagnostic further, first considering the
influence of the height on the strength of the Stokes $Q$ and $U$
profiles. Then, exploiting the sensitivity of the Hanle effect to the height
of the observed structure, we have analyzed again the data of May 13 2003, in
order to independently estimate the height to which these data refer to,
particularly at the location interpreted by \citet{solanki2003} to be the apex
of the highest reconstructed loop.

In Sect. \ref{obsdata} we summarize the data we analyze in this paper and introduce the codes we use to synthesize and invert the Stokes profiles. In
Sect. \ref{results} we present results of a parameter study of the strength of
the linear polarization signal. We also compare the observed with the
theoretical profiles in order to fix a minimum height and perform inversions
with \helixp{}, which allow the field strength, direction and the height to be
determined.

\section{Observational data and codes \label{obsdata}}

\subsection{Data}

Spectropolarimetric maps of the emerging flux region NOAA 9451 were recorded
on 13 May 2001 using the Tenerife Infrared Polarimeter
\citep[TIP,][]{martinezpillet1999}. The covered wavelength range contains the
10830 \AA\ multiplet. The active region was located near disk center at $\mu =
{\rm cos} \Theta = 0.8$. These observations have been analyzed in several
papers \citep{solanki2003,lagg2004,lagg2007} which emphasize different aspects
of this rich data set. In the present paper we focus on one spatial pixel of
the observed map situated in the central part of the active region \cite[pixel
$x=72, y=43$ of Fig. 1 in ][]{lagg2004}.

 The Stokes profiles at this pixel are also displayed in Fig. 3 of \cite{lagg2004}. This point is within the area of few pixels where
the Hanle effect in forward scattering is most prominent in the Stokes
profiles. Around this area the linear polarization rapidly falls below the noise level. This pixel also displays the largest
linear polarization signal in the entire map and, importantly, it corresponds to the apex of the highest loop
reconstructed by \citet{solanki2003}. We refer to this pixel
as the \textsl{loop apex}.  It makes sense to concentrate on this loop-top
pixel since it displays the largest discrepancy between the interpretations of
\citet{solanki2003} and \citet{judge2009} and the large linear polarization alsomakes this pixel the best location to demonstrate the new technique.

The degree of linear polarization reaches $0.45 \%$ of the continuum intensity in
the center of the red component of the multiplet \citep[see Fig. 3
of][]{lagg2004}.  The observed polarization is approximately parallel to the
dark lines visible in the core of the red \hei{} component which connect the
footpoints of the reconstructed loop spanning the emerging flux region. These
elongated absorption structures \citep[see Fig. 1 of][]{solanki2003,lagg2004}
are reminiscent of $H{\alpha}$ arch filament systems. The Stokes $V$ signal in
this region is comparable to the noise level of 10$^{-3}$ of the continuum
intensity. The surrounding pixels show similar Stokes spectra with slightly
lower linear polarization signal, but enhanced circular polarization signal.

\subsection{Synthesis code \label{syncode}}

The Hanle profiles observed on disk are produced by the presence
of atomic polarization (i.e. population imbalances and quantum
coherences between the magnetic sublevels pertaining to the upper
and lower term of the 10830 \AA\ multiplet; i.e. ${}^3 P_{2,1,0}$
and ${}^3 S_{1}$ respectively), generated proportionally to the
degree of anisotropy of the unpolarized radiation field coming
from the underlying layers of the solar atmosphere. This atomic
polarization, modified in the presence of a magnetic field  by the
Hanle effect in forward scattering \citep{trujillobueno2002},
selectively absorbs and re-emits the unpolarized light coming from
solar disk center, acting as sinks and sources of linear
polarization.

To reproduce these observed Hanle profiles we use a synthesis code with the
same forward calculation module as described in \cite{landi1982},
\cite{landi2004} and \cite{asensioramos2008}.

The Helium absorption is modeled to take place in a constant-property slab
located at a given height $h$ above the solar surface and illuminated from the
underlying solar photosphere. In this situation the degree of atomic
polarization of the Helium atoms is a function of (1) the anisotropic
illumination that produces this population imbalance, and (2) the presence of
a magnetic field that modifies it. In a plane parallel atmosphere the
anisotropy of the radiation at a given $\lambda$ depends basically on the
height of the scattering atom.

In order to quantify the atomic polarization we solve the statistical
equilibrium equations for the multipole components $\rho^K_Q (J,J')$ of the
atomic density matrix for a multi-term atom. We take into account coherences
between different $J$-levels \citep[see chapter 7 in][]{landi2004}, the
anisotropy of the radiation field for a given height \citep[computed for the
center-to-limb variation tabulated in][]{pierce2000} and the presence of a
magnetic field. Note that the sophistication with which the polarization
produced by the Hanle effect is computed here is considerably greater than in
the analysis of \citet{solanki2003}. From the calculated density matrix
elements it is possible to compute the emission coefficients $(\epsilon_I,
\epsilon_X$ with $X = Q, U, V)$ and the absorption coefficients ($\eta_I,
\eta_X$) for polarized light. To calculate the emergent Stokes parameters we
use the solution of the radiative equations for a slab of constant properties
\citep[see, e.g.,][]{asensioramos2008} taken to lie in the upper chromosphere

\begin{eqnarray}
I(\tau) &=& I_0 e^{-\tau} + \frac{\epsilon_I}{\eta_I}(1 -
e^{-\tau}), \\
 X(\tau) &=& X_0 e^{-\tau} + \frac{\epsilon_X}{\eta_I}(1 - e^{-\tau}) -
  \frac{\epsilon_I \eta_X}{\eta^2_I}(1 - e^{-\tau})\\
        &+& \frac{\eta_X}{\eta_I}\tau e^{-\tau}
  \Big(\frac{\epsilon_I}{\eta_I}-I_o\Big), \nonumber
\end{eqnarray}

\noindent where $I_0$ and $X_0$ are the boundary conditions
representing the radiation entering the slab from below. We take
$I_0$ to be the photospheric continuum at $\mu=0.8$ as tabulated
by \cite{pierce2000} and $X_0 = 0$. This solution differs from the
Milne-Eddington solution employed in earlier analysis of the
observed Stokes profiles considered here. The slab geometry has
the advantage that it allows a unique height to be assigned to the
center of the absorption region. It may also be closer to the case
of \hei{} formation in an emerging flux region.

\subsection{Inversion code \helixp{} \label{invcode}}

The forward model described in the previous section is implemented in the
inversion code \helixp{} \cite[]{lagg2010}. The free parameters of the
constant-property slab model are the magnetic field strength, inclination and
azimuth, $B$, $\gamma$ and $\chi$ (in the solar reference frame), the line of
sight velocity, $v_{LOS}$, the damping parameter $a$, the thermal broadening
of the line, $v_{Dopp}$, the optical thickness of the slab, $d$, and the
height of the slab above the solar surface, $h$. Following the analysis of
\cite{solanki2003} we assume the filling factor to be unity. The Pikaia
genetic algorithm \cite[]{charbonneau1995} was used to maximize the fitness
function $f$, defined as the reciprocal of the squared difference between the
measured and the observed Stokes vector $P$ ($P=I,~Q,~U,~\mbox{or}~V$) over a
wavelength window of ($-1.8$,+0.9) \AA\ centered at the red lines of the
\hei{} triplet, divided by the strength $s$ of the observed profiles
($s_I=\sum\limits_\lambda |I(\lambda)-1|$ and $s_{Q,U,V}=\sum\limits_\lambda
|(Q,U,V)(\lambda)|$) and the number of free parameters $N_{free}$:

\begin{equation}
  \frac{1}{f} = \sum\limits_{P}\frac{1}{N_{free}} \sum\limits_\lambda \frac{w_P(\lambda)}{s_P} \,
  [P_{fit}(\lambda) -  P_{obs}(\lambda)]^2
\end{equation}

The weighting functions $w_P$ were chosen to be unity within $\pm$0.6 \AA\
around the core of the red component of the \hei{} triplet, and was set to 0.6
outside this range.  To take into account the reduction of the intensity level
caused by the wings of the photospheric Si~\textsc{I} at 10827 \AA\ we fit
this line with a Voigt function. The parameters of this Voigt function are
determined from an average of 100 Stokes $I$ profiles centered around the loop
apex.

In extensive tests we found the Pikaia algorithm to be the most reliable
technique to determine the global minimum within the considered parameter
space independently of the choice of the initial guess values. Additionally,
the random walk convergence of the Pikaia algorithm allows for a comprehensive
error analysis in a statistical sense. The fluctuation of the retrieved
parameters for individual, independent inversions of the same Stokes vector
naturally delivers a measure for the reliability and robustness of the
retrieved parameters.

\section{Results \label{results}}

We have performed computations with both codes introduced in Sect.
\ref{syncode} and \ref{invcode}. These computations have different aims. In a
first step, described in Sect. \ref{hdep}, we obtain an idea of the dependence
of the linear polarization in the \hei{} triplet on the components of the
magnetic vector and the height of the slab. Then, in Sect. \ref{compsynth} we
test whether this technique can be employed to set lower or upper limits on
the height of the scattering atoms by comparing with a given observed
profile. Finally, in Sect. \ref{invresult}, we obtain a best estimate of the
height of the sampled field using an inversion technique.

\subsection{Dependence of the computed polarization on height and magnetic
  field \label{hdep}}

\begin{figure*}
    \includegraphics[width=\linewidth,clip=TRUE]{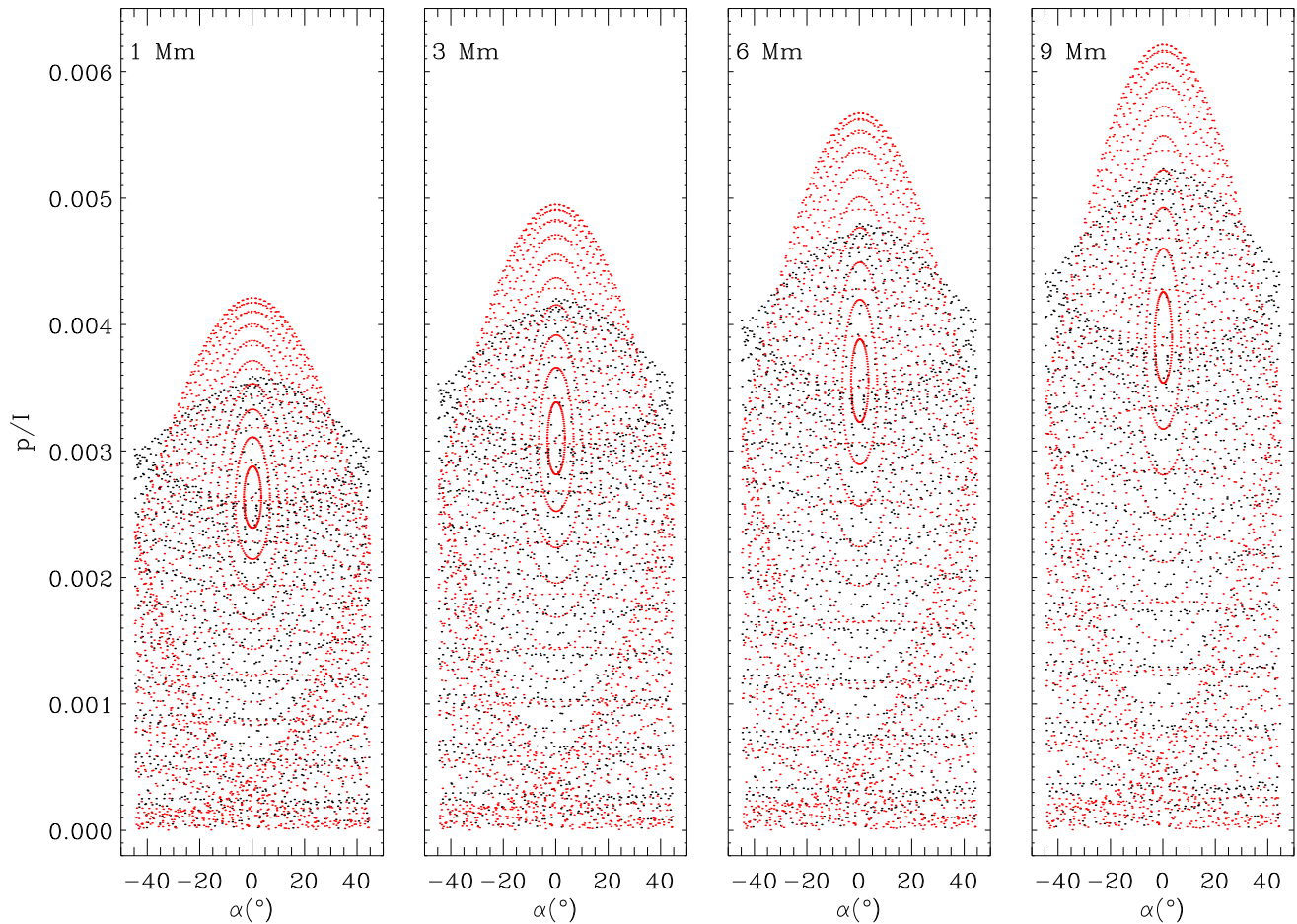}
    \caption{Polarization diagrams computed for different heights above the
      solar surface (from 1\,Mm in the leftmost frame to 9\,Mm in the rightmost
      frame). In every diagram, the y-axis gives the degree of linear
      polarization $p=\sqrt{Q^2+U^2}/I$, the x-axis the polarization
      angle $\alpha = 0.5 \, {\rm tan}^{-1}(U/Q)$. $p$ represents the maximum
      in the wavelength range covered by the \hei{} multiplet. The plotted
      $\alpha$ has been sampled at the same wavelength. Red points correspond
      to ``vertical'' magnetic fields ($\theta_B < 54.7^{\circ}$ or $\theta_B
      > 125.3^{\circ} $), while black points correspond to ``horizontal''
      magnetic fields ($ 54.7^{\circ} < \theta_B < 125.3^{\circ}$). }
    \label{fig:diagrams}
 \end{figure*}

The way in which the presence of a magnetic field modifies the
emitted polarization signal depends on its direction and on the
position of the observed point on the solar disk. For example,
off-limb linear polarization emission is maximum in the absence of
a magnetic field, or for a vertical magnetic field (with respect
to the solar surface). Its value decreases for increasingly
horizontal magnetic fields. For observations at solar disk center,
on the contrary, the He atoms do not emit linear polarization in
the absence of a magnetic field or for a vertical magnetic field.
Here the intensity of linearly polarized radiation increases for
increasingly horizontal magnetic fields. For observations at other
locations of the solar disk a behavior somewhere in between these
two extremes is present. Figure \ref{fig:diagrams} displays the
results of an extensive parameter survey, which aims to determine
the dependence of properties of the linear polarization on the
parameters of the magnetic vector and the height of the slab. The
computed examples refer to $\mu = 0.8$, which corresponds to the
location of NOAA 9451 at the time it was observed in \hei{}. Each
dot represents the degree of linear polarization
$p=\frac{\sqrt{Q^2+U^2}}{I}$ plotted versus the polarization angle
$\alpha = 0.5 \, {\rm tan}^{-1}(U/Q)$ (both at the $\lambda$ of
maximum $p$) for profiles computed for all possible orientations
of the magnetic field. The inclination angle with respect to the
local solar vertical, $\theta_B$, is varied from $0^{\circ}$ to
$180^{\circ}$ in steps of $2^{\circ}$ ($\theta_B = 0^{\circ}$ for
a magnetic field vector pointing outward from the Sun) and the
azimuth angle in the plane parallel to the solar surface,
$\chi_B$, is varied from $-90^{\circ}$ to $90^{\circ}$ in steps of
$4^{\circ}$ ($\chi_B = 0^{\circ}$ for a magnetic field vector
lying in the plane of the local solar vertical and the observer
direction). At every distance above the solar surface there is a
maximum polarization, corresponding to ``vertical'' magnetic
fields (less inclined than Van Vleck's angle $54.7^{\circ}$ to the
solar surface normal, red points in the figure). This maximum
increases with height. The same increase occurs for the
``horizontal'' field (more inclined than Van Vleck's angle, black
points in the figure), although in this case the maximum reached
is lower. Note that irrespective of the height there is no minimum
polarization. This behavior implies that for a given measured
linear polarization magnitude a lower limit on the height of
scattering atoms can be placed, but no upper limit (just based on
the degree of linear polarization). Furthermore, this lower limit
is higher for a horizontal magnetic field than for a vertical
field.

\subsection{Comparing observed with synthetic profiles for
different heights \label{compsynth}}

\begin{figure*}
    \includegraphics[width=\linewidth,clip=TRUE]{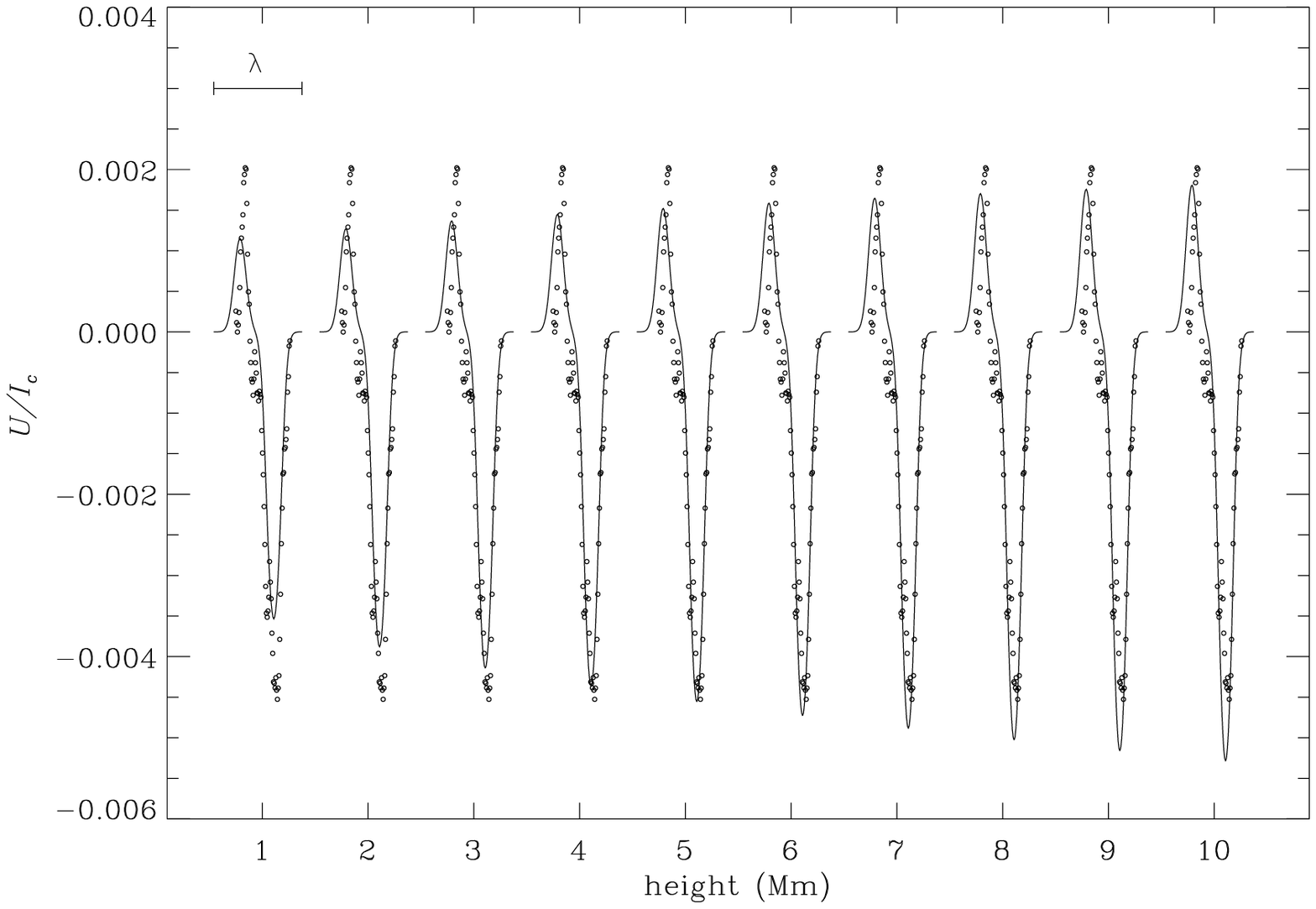}
    \caption{Observed Stokes $U/I_c$ profile (points) of Fig. 3 of
      \citet{lagg2004} compared with the maximum $U/I_c$ profiles computed for
      each height (solid line). At each height the observed and computed
      profile is plotted vs. $\lambda$.  The same observed profile is repeated
      for comparison at each height. The computed Stokes $I$ profile (not
      shown here) fits the observed intensity. The reference direction for
      positive Stokes $Q$ forms an angle $12^{\circ}$ with the nearest solar
      limb. This reference direction gives $Q/I_c\approx0$. The horizontal bar
      indicates the wavelength range, 10827.9 - 10831.6 \AA\, of the leftmost
      profile.}
    \label{fig:maxima}
 \end{figure*}

In Fig. \ref{fig:maxima} we compare the linear polarization profile of Fig. 3
of \citet{lagg2004}, pixel x$=72$, y$=43$ in their Fig. 1, (points) with the computed profile displaying the maximum
linear polarization for a set of heights between 1 and 10\,Mm above the solar
surface. Each solid profile is plotted versus wavelength and corresponds to
the height given on the horizontal axis. Each synthetic profile is the one
showing the largest $U/I_c$ among all profiles in the ``horizontal'' range,
i.e., more inclined to the vertical than $54.7^{\circ}$, formed at that height
(for both synthetic and observed profile the reference direction for positive
Stokes $Q$ form an angle of $12^{\circ}$ with the nearest solar limb,
resulting in $Q/I_c \approx 0$). The requirement that the magnetic field is in
the ``horizontal'' range is driven by the fact that the Stokes $V$ signal is
comparable to the noise level.  Additionally, the Stokes $V$ spectra along the
loop structure change polarity close to the loop apex. It is also consistent
with the finding of \citet{solanki2003} that this location corresponds to the
top of a loop. Clearly, for heights lower than 5\,Mm the anisotropy of the
radiation field that excites the Helium atoms is not able to produce
sufficient atomic polarization to reproduce the observed $U$ profile. In particular, a slab
at a constant height of 2.4\,Mm, as proposed by \citet{judge2009}, does not
provide a satisfactory description, since even the strongest $U$ profile from
that slab fails to achieve the polarization level of the observed profile.  It
falls well short of the polarization in the 5 most strongly polarized points
of the observed profile.

\subsection{Results of inversions \label{invresult}}

We carried out a series of inversions (see Section \ref{invcode} for a
description of the code and setup) in order to determine the atmospheric
parameters which best describe the observed Stokes profiles shown in Fig. 3 of
\citet{lagg2004}.  The minimization technique used for the inversions with
\helixp{} involves the genetic algorithm PIKAIA \citep{charbonneau1995}. This
algorithm has the advantage of finding the global minimum (i.e. the synthetic
Stokes profile best matching the observed one) within the considered parameter
space completely independent of any initial guess values. However, the path of
convergence to this global minimum is random and may require an infinite
number of iterations. To overcome this problem we repeated the inversions 100
times, each with a sufficiently high number of iterations ($n=500$). The
result of the inversion run with the best fitness out of the 100 independent
runs was considered to be the solution closest to the global minimum.

\begin{figure}
  \centering
  \includegraphics[width=1.\linewidth,clip=TRUE]{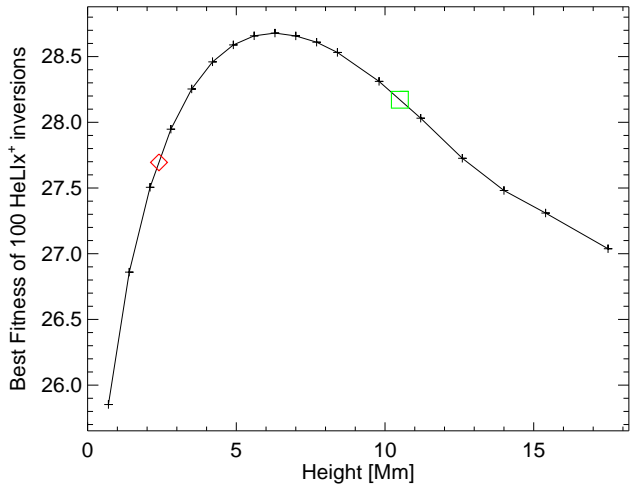}
  \caption{Best fitness out of 100 individual inversion runs for fixed height
    values from 0.7\,Mm to 17.5\,Mm. The optimum fit to the observed Stokes
    vector is achieved for a height of the He slab of
    $\approx$6.3\,Mm. The diamond and the rectangular symbol indicate the
      fitness at a height of 2.4\,Mm and 10.5\,Mm.}
  \label{bestfit100}
\end{figure}

The allowed range for the values of the the atmospheric parameters for the
individual inversion runs is presented in Tab.  \ref{tab:range}. These ranges
represent a carefully selected compromise between, on the one hand, stability
and speed of the inversions and, on the other hand, the full coverage of
expected values for the atmospheric parameters at the loop top.  We carried
out 100 independent inversion runs on the loop top pixel for each fixed He
slab height ranging from 0.7\,Mm to 17.5\,Mm. The best fitness of the 100
inversions at each height is plotted as a function of height in
Fig. \ref{bestfit100}. The figure clearly shows that a global maximum of the
fitness is achieved for a height of the He slab at the loop top of
$\approx$6.3\,Mm. Note that a higher fitness corresponds to a better fit.  The
smoothness of the maximum fitness curve in Fig. \ref{bestfit100} indicates
that even relatively small fitness differences are significant.  The fitness
at the height of 2.4\,Mm, proposed by \citet{judge2009} is marked in
Fig. \ref{bestfit100} by a diamond. It is significantly lower than the best
fitness. Additionally we marked the fitness corresponding to height of
10.5\,Mm as retrieved by \cite{solanki2003}. 

\begin{figure}
  \centering
  \includegraphics[width=1.\linewidth,clip=TRUE]{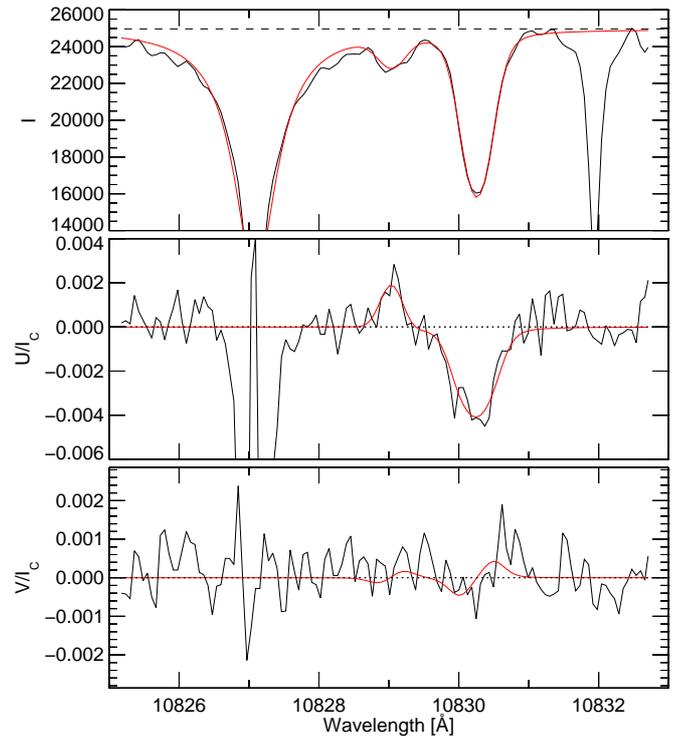}
  \caption{The observed Stokes profile analyzed in this paper (solid black
    curves; left vertical scale; same reference direction as
    Fig. \ref{fig:maxima}).  The horizontal dashed line in the top panel corresponds to the
    adopted continuum level. Also plotted (red solid lines) are best fit
    profiles for a height of 6.3\,Mm, obtained for a magnetic field of 320 G,
    with an inclination of $95^{\circ}$ with respect to the local solar
    vertical, and an azimuth of $-86^{\circ}$ (see Section \ref{invresult}).}
  \label{fig:bestfit}
\end{figure}

Fig. \ref{fig:bestfit} shows the fit obtained from the best
inversion run for a slab centered on a height of 6.3\,Mm. Lower
height values fail to
  produce a sufficiently strong linear polarization signal. If the absorbing
  Helium slab is located higher than 6.3\,Mm, the linear polarization signal
  resulting from scattering would be larger than the observed one. The
  inversion procedure compensates these strong signal by increasing the
  magnetic field strength, since a stronger magnetic field leads to
  depolarization of the scattering polarization. However, with increasing
  magnetic field strength the Zeeman signal increases, resulting in a mismatch
  for the Stokes $V$ signal and consequently in a reduction of the fitness.
The atmospheric parameters from the best inversion run are displayed in the
fourth column of Tab. \ref{tab:range}. The magnetic field strength at the loop
top is 311\,G, the field is nearly parallel to the solar surface (inclination
angle $\gamma$=95$^\circ$), and the azimuth angle confirms the orientation of
the field along the visible structures in the He equivalent width image in
Fig. 1 of \citet{solanki2003}.  The up-flow of 1.3\,km/s at the loop top
confirms the rising flux tube scenario presented in that paper.

The simpler analysis of \cite{solanki2003} provided a field strength that
dropped to around 50 G at the loop apex, while here we obtain 320 G. The
higher value lies considerably closer to the field strength that
\citet{wiegelmann2005} retrieved at the apex of the loops closest to
the one whose apex we study in this paper using a force-free extrapolation
from the photospheric magnetic field map (305 - 615 G). This illustrates the
importance of using the more realistic treatment of the Hanle effect employed
here compared to that employed by \citet{solanki2003} and \citet{lagg2004}.

We also have inverted the few pixels showing linear polarization located near this one, finding comparable values of magnetic field and  slightly lower heights, consistently with the fact that the pixel discussed here in detail was located at the \textsl{loop apex}.

\begin{table}[h]
  \caption{Ranges for the atmospheric parameters. The fourth column shows the atmospheric parameters for the best fit.}
\label{tab:range}
\begin{tabular}{l c c c}\hline
parameter           & min.         &  max.       & best fit  \\
                    & value        &  value      & ($h=6.3$\,Mm) \\
\hline
$B$ [G]             & 5            & 1000        & 311\\
$\gamma$ [$^\circ$] & 60           & 120         & 95\\
$\chi$ [$^\circ$]   & -90          & 90          & -86\\
$v_{LOS}$ [km/s]    & -10          & 10          & -1.3\\
$a$                 & 0.05         & 0.50        & 0.22\\
$v_{Dopp}$  [km/s]  & 5            & 25          & 6.80\\
$d$                 & 0.20         & 0.90        & 0.47\\
\hline
\end{tabular}
\end{table}

\section{Discussion and conclusions \label{discussion}}

We have shown that an analysis of the linear polarization of the \hei{} 10830
\AA\ triplet using an atomic polarization interpretation can be used to
estimate the height at which the triplet is formed and hence the height at
which it samples the local magnetic and velocity field. Lower limits
on this height are  more straightforward to obtain than upper limits. Because the lower limit is tighter for larger polarizations, this technique works best for
profiles displaying large-amplitude Hanle-like $Q$ or $U$ signals.

\citet{solanki2003} have proposed a reconstruction technique to obtain the 3D
magnetic field structure of an emerging flux region observed near the disk
center.  This technique is based on the assumption that the He
10830 \AA\ line is formed along the magnetic flux tube as it emerges through
the chromosphere into the corona. The reconstruction was constrained by
requiring the physical quantities in the loops to be smoothly varying and for
both footpoints of the loops to lie at the same height. They found that the
tallest reconstructed loop reached in its central part a height of $\approx
10$\,Mm over the solar surface. Recently \citet{judge2009} has proposed an
alternative interpretation of this observation. He argues that the
observed magnetic fields are located in a thin layer at a constant height
($\approx 2.4$\,Mm) at the top of the basically plane-parallel chromosphere.

In the present paper we have analyzed the most critical pixel in the
observation of \cite{solanki2003}, namely that at the apex of the highest
loop, with the help of \helixp{}, an inversion code incorporating both Hanle
and Zeeman effects. We found the best inversion fit for an height of $\approx
6.3$\,Mm. This value is lower that the value found by \citet{solanki2003}, but
the inversions favor the interpretation of the 3D loop reconstruction of
\citet{solanki2003} over the interpretation proposed by \citet{judge2009} because the fitness of the results (a measure of the goodness of the fit)
displays an asymmetric behavior, dropping rapidly towards lower heights, but
comparatively gently towards greater heights.

Further support for the above conclusion is provided by an
inversion using \helixp{} of a profile at the apex of emerging
loops in AR NOAA 10917 presented by \citet{xu2009}. For that
active region the inversion including atomic polarization returns
a height of 7.1$\pm$2\arcsec{} ($5.0\pm 1.4$\,Mm), which is not
only higher than the formation height of 2.4\,Mm favored by
\citet{judge2009}, but also higher than the reconstructed looptops
at 4\,Mm. Consequently, there is some discrepancy between the
heights obtained from magnetic field reconstructions and from the
Hanle effect, but in the two cases studied so far the difference
has opposite sign, which suggest that it is due to statistical
errors. From the present analysis we cannot judge if the error is
larger in the heights of the reconstructed loops or in the heights
deduced from the Hanle effect. In both studied cases, however, the
height deduced from the Hanle effect is higher than that proposed
by \citet{judge2009}.

The interpretation put forward by \citet{judge2009} is probably valid in the
quiet Sun, where under ``normal conditions'' the He I 10830 \AA\ multiplet
tends to form in a thin layer at the top of the chromosphere. As Judge himself
pointed out, the He triplet is formed over ``a broad distribution of heights
owing to the corrugated nature of the chromosphere''. An example of a
structure in which \hei{} 10830 forms well above its quiet Sun formation level
are filaments or prominences.  Thus, in prominences He 10830 \AA\ has been
observed up to several tens of Mm above the limb.  The reason is the much
higher density of material at chromospheric temperatures inside
prominences and filaments than in the surrounding corona. The density is sufficient to make the line
optically thick. Our result suggests that the line forms at much greater
heights than usual in an EFR, just as it does in filaments. Young, recently
emerged loops in EFR, are often referred to as arch filament systems (AFS)
when observed in $H{\alpha}$ \citep[see, e.g., the review by][]{chou1993} due
to their arch-like shape and their high density of chromospheric material,
which leads to enhanced absorption, giving them a filament-like
appearance. The He triplet similarly displays enhanced absorption in both EFRs
studied to date, displaying dark filaments that look very similar to the ones
seen in $H{\alpha}$ \citep{xu2009}. These dark filaments lie parallel to the
magnetic field azimuth and are the locations at which loops can be
reconstructed. They do not show interruptions or inhomogeneities along their
long axes as one would expect in the interpretation favoured by Judge, but rather a smooth
change in brightness.  The chromospheric magnetic field changes equally
smoothly along these filaments, in stark contrast to the photospheric field,
which displays a very much higher level of inhomogeneity, including the
presence of opposite polarities below the loop tops (see Fig. 1 of
\citeauthor{solanki2003}, \citeyear{solanki2003}; Fig. 5 of
\citeauthor{xu2009}, \citeyear{xu2009}). Thus, the magnetic field structure in
an EFR is rather different from the simple potential field model used by
\citet{judge2009}, in which the photospheric and chromospheric fields are
basically identical.  These points also favor the interpretation of
\citet{solanki2003} and lend support to the reconstructed loops.

Finally, in his paper \citet{judge2009} did not put any emphasis on the fact that the 3-D structure of the reconstructed
loops agrees rather well with a non-linear force-free (NLFF) extrapolation
starting from a map of the photospheric vector magnetic field. This was
derived from an independent inversion of simultaneously recorded
spectropolarimetric measurements of the photospheric Si I 10827 \AA\
line. Since NLFF extrapolations have no free parameter and take into account
electric currents, this agreement is quite remarkable. Interestingly the work
of \citet{wiegelmann2005} also showed that a potential field is a very poor
representation of the chromospheric field, giving a completely different
azimuth. Hence, even if ignoring the problem that photospheric magnetic
structure is far more complex than the chromospheric one, it is unclear if
conclusions based on a (extremely simple) potential field computation
\citep{judge2009} have any validity for AR 9451.

In summary, a state-of-art treatment of the Hanle effect suggests
that the interpretation of \citet{solanki2003} is to be preferred
over that of \citet{judge2009}. More importantly, this paper has
demonstrated the value of treating the Hanle effect more
completely than the simple approximation employed by
\citet{solanki2003}. This more complete treatment leads not only
to the possibility of estimating the height of formation of the He
triplet, but also results in improved retrieved values of the
magnetic vector. The effect of the height on the line profiles is
rather small, however, and thus requires observations with
extremely low noise levels.

\acknowledgement{We thank E. Landi Degl'Innocenti for helpful
  discussions, and P. Judge for his comments on the manuscript. This work was supported by the EU through the EC FP7
  Marie-Curie European Reintegration Grant (contract no. MERG-CT-2007-211048,
  ``ChroMag''). This work was partly supported by the WCU grant No. R31-10016
  of the Korean Ministry of Education, Science and Technology. }

\bibliography{14988biblio}
\bibliographystyle{aa}

\end{document}